\documentclass[conference]{IEEEtran}
\ifCLASSINFOpdf
\else
\fi
\pdfoutput=1

\usepackage{epsfig}
\usepackage{algorithm, amsmath}
\usepackage[noend]{algpseudocode}
\usepackage{graphicx}
\usepackage{caption}
\usepackage{subcaption}
\usepackage{color}
\usepackage{cite}
\usepackage{multirow}
\usepackage{url}
\usepackage{comment}
\usepackage{wrapfig}
\usepackage{amsmath}
\usepackage{mathtools}
\usepackage{amssymb}
\usepackage{physics}
\usepackage{array}
\usepackage{nicefrac}
\usepackage[font=small,labelfont=bf]{caption}
\begin{document}
%
\title{Wireless Quantization Index Modulation: Enabling Communication Through Existing Signals}



%

%

\author{\IEEEauthorblockN{Zerina Kapetanovic\IEEEauthorrefmark{1},
Vamsi Talla\IEEEauthorrefmark{3},
Aaron Parks\IEEEauthorrefmark{1}, 
Jing Qian\IEEEauthorrefmark{2} and
Joshua R. Smith\IEEEauthorrefmark{1}\IEEEauthorrefmark{3}}
\IEEEauthorblockA{\IEEEauthorrefmark{1}Department of Electrical Engineering, University of Washington}
\IEEEauthorblockA{\IEEEauthorrefmark{2}Department of Electrical Engineering, Tsinghua University}
\IEEEauthorblockA{\IEEEauthorrefmark{3}Department of Computer Science and Engineering, Univeristy of Washington}}


\maketitle



%
\IEEEpeerreviewmaketitle

\makeatletter
\def\ps@IEEEtitlepagestyle{%
  \def\@oddfoot{\mycopyrightnotice}%
  \def\@evenfoot{}%
}
\def\mycopyrightnotice{%
  {\footnotesize 978-1-5386-1456-3/18/\textdollar31.00 ©2018 IEEE\hfill}
  \gdef\mycopyrightnotice{}
}

{\bf Abstract --} As the number of IoT devices continue to exponentially increase and saturate the wireless spectrum, there is a dire need for additional spectrum to support large networks of wireless devices. Over the past years, many promising solutions have been proposed but they all suffer from the drawback of new infrastructure costs, setup and maintenance, or are difficult to implement due to FCC regulations. In this paper, we propose a novel Wireless Quantization Index Modulation (QIM) technique which uses existing infrastructure to embed information into existing wireless signals to communicate with IoT devices with negligible impact on the original signal and zero spectrum overhead. We explore the design space for wireless QIM and evaluate the performance of embedding information in TV, FM and AM radio broadcast signals under different conditions. We demonstrate that we can embed messages at up to 8--200~kbps with negligible impact on the audio and video quality of the original FM, AM and TV signals respectively. 

\section{Introduction} \label{intro}

Over the last decade, we have witnessed a rapid growth in the deployment
of IoT devices. By some estimates, there will be more than 26 billion
connected IoT devices by the year 2020~\cite{iot2020}. However, as more
devices connect to wireless networks, \textit{\textbf{available spectrum
is insufficient and existing wireless protocols are ill-equipped to
support the growing number of devices.}} To understand the challenge,
consider a home with wireless cameras, security sensor, smart watches,
fitness trackers and wireless speakers. These devices use
Wi-Fi/Bluetooth or proprietary wireless in the 2.4~GHz ISM band and
operate alongside Wi-Fi routers, smartphones, laptops and tablets. As
more devices share the wireless channel, wireless interference and
packet collisions increase, negatively impacting the throughput and
latency~\cite{musaloiu2008minimising}~\cite{gollakota2011clearing}.

As wireless spectrum (such as the 2.4~GHz ISM band) becomes crowded,
conventional wisdom dictates that we migrate to new protocols in less
congested wireless channels. New protocols such as 802.11ah,
LoRaWAN~\cite{lorawan}, SIGFOX~\cite{sigfox} operate in 915~MHz ISM
band, high speed 802.11 n/ac Wi-Fi is moving to 5.8~GHz ISM band, NB-IoT
operates in the licensed cellular bands and TV white space
networking~\cite{whitespace} operates in unused channels in the TV UHF
spectrum. Although these solutions are a step in the right direction,
let's discuss these approaches in terms of cost and spectrum
utilization.

\begin{figure}[t] \vspace{-0.05in} \centering
\includegraphics[width=0.8\columnwidth]{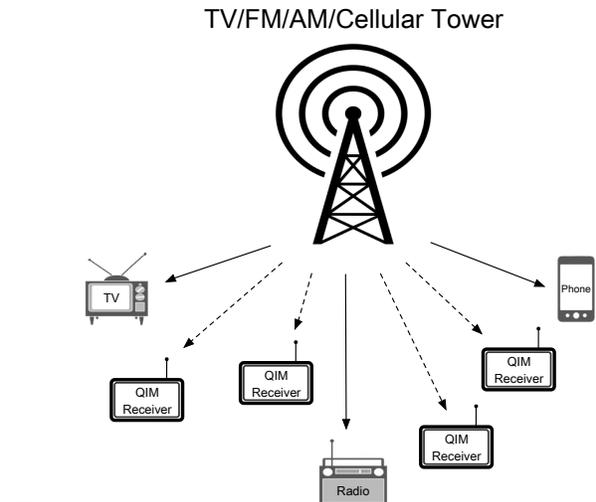}
\vspace*{-0.15in}\caption{\footnotesize{\textbf{QIM deployment.} QIM
uses existing wireless signals to communicate with IoT devices.}}
\label{fig:model} \vspace{-0.3in} \end{figure}

\begin{itemize}

\item \textit{Infrastructure and Maintenance Costs}: Migration to new
protocols such as LoRaWAN, SIGFOX or 802.11ah require setup, deployment
and maintenance of dedicated expensive gateways and base stations.
Instead, if we can reuse existing wireless infrastructure for
communication, we could develop a far simpler and cost effective
solution.

\item \textit{Spectrum Utilization}: Wireless spectrum is an extremely
valuable and highly regulated resource. TV white space networking uses
allocated but otherwise under-utilized TV spectrum for wireless
communication. However, more often than not the availability of unused
TV channels in urban areas are scarce and it is very cumbersome and
expensive to deploy a TV whitespace network~\cite{fcc}. New protocols
such as LoRaWAN, SIGFOX, and 802.11ah are moving to the less crowded
915~MHz ISM band, but over time as the number of devices increase, they
are going to run into familiar interference and capacity issues: as the
number of devices increase, the spectrum is going to become more crowded
and eventually saturate. Unless new spectrum is made available, using
traditional methods, it is impossible to scale beyond a certain point.

%
\end{itemize}

In this paper, we propose a new cost and spectrally efficient solution
for wireless communication. Consider an urban city environment as shown
in Fig.~\ref{fig:model} with existing deployments of AM, FM, TV, and
cellular base stations. These base stations have been setup with
tremendous infrastructure cost, undergo periodic maintenance and pay
licensing fees to transmit at pre-assigned licensed frequencies. The
base stations are designed and geographically located for optimal signal
coverage. For example, a typical FM tower can be received up to 100~kms.

We introduce Wireless Quantization Index Modulation (QIM), a
communication technique which leverages existing infrastructure and
reuses broadcast signals to provide additional communication channels
for IoT devices. To understand Wireless QIM, without the loss of
generality, let's consider a broadcast TV station. A TV transmitter can
use the QIM technique in its baseband to embed a message into the
broadcast TV signal by introducing small perturbations while having a
negligible impact on the broadcasted TV signal. Legacy TV receivers in
the coverage area decode the broadcasted signal as before while IoT
devices with a QIM receiver can decode the embedded message without any
prior knowledge of the broadcast signal. So, in summary with a small
modification to the baseband of the broadcast station, Wireless QIM reuses infrastructure, spectrum and broadcast signals to
simultaneously communicate with QIM enabled IoT devices and legacy
AM/FM/TV/cellular devices.

Wireless communication requires both uplink and downlink. However, more
often than not, it's asymmetric i.e. depending on the application,
either uplink or downlink communication dominates. In this paper, we
focus on downlink heavy applications and design a Wireless QIM system
for downlink communication. Our target application is a smart city where
using Wireless QIM, existing wireless infrastructure provides
connectivity for real-time update of electronic bus schedule displays,
billboard signs and advertisements, traffic alerts to name a few. With a
minimal change in the baseband of existing broadcast towers, we can
embed data to wirelessly update devices with a QIM receiver in
real-time. These applications would require a minimal uplink channel to
send acknowledgement messages, however such a low bandwidth and
infrequent task can be accomplished using traditional LoRa, SigFox or
cellular radios for the time being. In future work, we will extend the
Wireless QIM technique to uplink communication and develop a bi-directional
communication system which can leverage existing infrastructure and
communicate with smart devices with zero spectrum overhead and minimal
additional cost to target a broader set of applications.

To demonstrate the efficacy of Wireless QIM for these applications, we
extensively evaluate the design space and explore various tradeoffs
between performance of the message and host signal. We implement Wireless QIM on
three existing infrastructure broadcast signals: AM, FM, and TV and show
that information can be reliably embedded with negligible impact on
audio (AM and FM) and video (TV) quality of the host signals. Our results show that using Wireless QIM, we can embed messages for IoT devices at up
to 8~kbps in AM radio signals and 200~kbps in FM signals.
%


\begin{figure}[t] \vspace{-0.05in} \centering
\includegraphics[width=1\columnwidth]{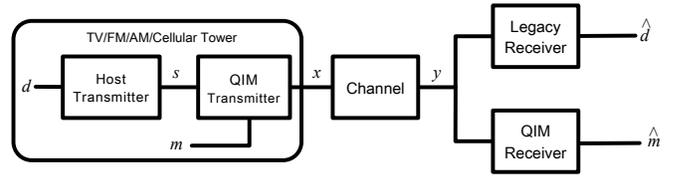}
\vspace*{-0.15in} \caption{\footnotesize{QIM embedding model, where a
message $m$ is embedded in a host signal $s$ using the QIM embedding
function. The signal passes through a noisy channel and the decoder
retrieves the estimated message $\hat{m}$ from the received signal, $y$.
}} \label{fig:qim_blk} \vspace{-20pt} \end{figure}

\section{Quantization Index Modulation} \label{sec:methods} Quantization
Index Modulation (QIM) was originally introduced as a scheme for 
information hiding and digital watermarking~\cite{qim}. In these
applications, a message signal is embedded inside another signal called
the host signal such that the embedded message is robust to common
degradations, while the host signal suffers minimal degradation. QIM
technique achieves an efficient tradeoff between the data
rate of the message signal, distortion, and robustness of embedding.

In this work, we use QIM for wireless communication. We show how to
embed messages for IoT devices inside existing wireless signals at
existing wireless transmitters such that IoT devices with QIM receivers
can decode the messages while the broadcasted wireless signal
experiences minimal degradation. Fig.~\ref{fig:qim_blk} shows the block
diagram for a typical Wireless QIM system. At the AM/FM/TV/cellular
broadcast, data $d$ needs to be transmitted and is passed
through the host transmitter to generate a baseband host signal $s$.
The host signal can be of any form, for instance, a frequency modulated
signal from a FM station. Now, we also want to embed a message $m$
intended for IoT devices. The host signal is passed through a baseband
QIM transmitter that uses a quantizer $Q(s)$ to embed the message inside
the host signal. This generates a composite signal $x$ that propagates
through the (noisy) channel and signal $y$ is received at the receiver. A
legacy device would use standard compliant host receiver to decode the
host signal, while an IoT device would use a QIM receiver to decode the
embedded message $m$. QIM is designed to ensure that both the IoT and
legacy devices can decode the desired signals with minimal degradation.


A Wireless QIM system consists of a QIM transmitter at the base station
and a QIM receiver at the IoT device. To understand how Wireless QIM
works, we will first describe the QIM transmitter followed by the QIM
receiver.

\begin{figure}[th] \vspace{-12pt} \centering
\includegraphics[width=0.5\textwidth]{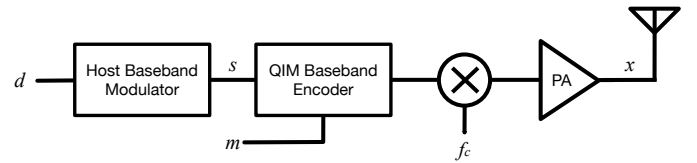}
\vspace*{-0.15in}\caption{\footnotesize{Block diagram of a QIM enabled transmitter.}}
\label{fig:qim_tx} \vspace{-10pt} \end{figure}

\subsection{QIM Transmitter} \label{sec:scalar}

On a high level, a QIM transmitter embeds information by using a form of quantization to introduce small perturbations in the host signal. Fig.~\ref{fig:qim_tx} shows the block diagram of a QIM transmitter, where data $d$ is passed through the baseband host modulator to generate the host signal $s$. Then the host signal and message $m$ are passed through the QIM encoder to generate a composite signal $x$ in baseband. Finally, the composite signal is upconverted to carrier frequency $f_c$ and transmitted. The implementation of the baseband QIM encoder depends on the host signal. For analog systems such as AM radio, QIM encoder consists of the digital baseband followed by the ADC, a standard component. Whereas in digital FM radio and TV systems, the QIM encoder can be integrated into pre-existing digital baseband. In the following section we explain how the QIM embedding process works. First lets define a uniform quantizer $Q\left(s\right)$ as


\begin{equation}
\vspace{-2pt}
Q(s) = \Delta\cdot
round(\frac{s}{\Delta})
\vspace{-2pt}
\label{eq:quantize} 
 \end{equation}

where $\Delta$ is the quantization step size and is defined as

\begin{equation}\label{delta} 
\vspace{-2pt}
\Delta = \frac{2\cdot max(s)}{N}
\vspace{-2pt}
\end{equation}

Here $N$ is the number of quantization levels. The step size and number of levels determine the embedding resolution for the host signal. Quantizer \emph{Q(s)} can now be used to define the QIM embedding function,

\begin{equation} \label{eq:qim} 
\vspace{-2pt}
Q_m(s) = Q(s - d_m) + d_m
\vspace{-2pt}
\end{equation}

where $d_m$ is the dither, a function of the message $m$ that is applied
to the host signal. $d_m$ can take one of the two following values to
represent embedding of either a 0-bit or 1-bit. 

\begin{equation} 
\vspace{-2pt}
\label{dither} d_1=\pm\frac{\Delta}{4} \text{ and } d_0=\begin{cases} d_1 + \frac{\Delta}{2}, & \text{if } d_1 \leq 0\\ d_1 -
\frac{\Delta}{2}, & \text{otherwise} \end{cases} 
\vspace{-2pt}
\end{equation}

The equation shows that if the 1-bit dither is negative, $d_1$ =
$\nicefrac{-\Delta}{4}$, then 0-bit dither, $d_0$ =
$\nicefrac{\Delta}{4}$, will be a positive value. Similarly, for a
positive 1-bit dither, the 0-bit dither would be a negative value. So, in summary we create two dithered quantizers to embed data in the host signal which is dependent on the bit value of the embedded message $m$.

\begin{figure}[th] \vspace{-15pt} \centering
\includegraphics[width=0.5\textwidth]{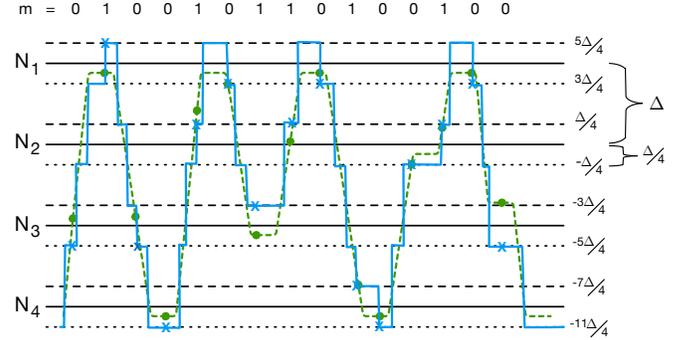}
\vspace*{-20pt}\caption{\footnotesize{Embedding using QIM for $d_1 =
\frac{\Delta}{4}$ and  $d_0 = \frac{-\Delta}{4}$}}
\label{fig:scalar_basic} \vspace{-20pt} \end{figure}

Fig.~\ref{fig:scalar_basic} illustrates this QIM
embedding process for a
positive 1-bit dither $d_1$ with $N=4$ levels, represented by solid horizontal lines. The dashed horizontal lines represent quantization levels for the 1-bit dithered quantizer and dotted lines for the 0-bit quantizer, defined by Eq.~\ref{eq:qim}. We can see that because the dither for a 1-bit is defined as $\nicefrac{\Delta}{4}$, the dashed lines are shifted up by $\nicefrac{\Delta}{4}$ from the original set of levels and vise versa for 0-bit quantizer. 

Each sample point of the host signal (dashed green line) is perturbed to the appropriate level depending on the message $m$ (shown at the top of the figure). For instance, at the first highlighted sample point (green dot) we encode a 0-bit and the composite signal (solid blue line) goes down to $\nicefrac{-5\Delta}{4}$, the nearest 0-bit level (blue X). Similarly, at the second highlighted point we embed a 1-bit, which jumps up to $\nicefrac{5\Delta}{4}$, the nearest level for a 1-bit. We can see from the example that the distance between quantization points is uniformly distributed between [$\nicefrac{\Delta}{2}$, $\nicefrac{-\Delta}{2}$]. As a result, the mean error due to embedding is equal to $\nicefrac{\Delta^2}{12}$. 



Finally, we characterize the impact of the QIM encoder on the host
signal. We define distortion in the host signal by comparing the
original host signal to the composite signal generated by embedding and
can be expressed as,

\begin{equation}\label{eq:distortion} \vspace{-2pt}
D_s = \frac{1}{K} \sum_{i=1}^{K}
|s_i - x_i|^2 
\vspace{-2pt}
\end{equation}

%

\bigskip

\begin{figure}[th] \vspace{-10pt} \centering
\includegraphics[width=0.5\textwidth]{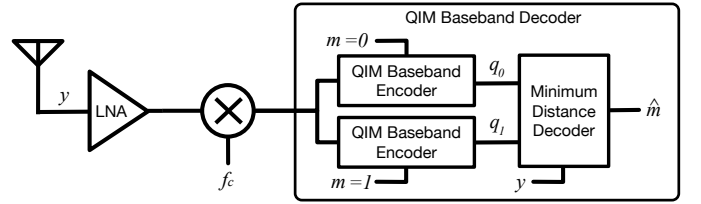}
\vspace*{-0.15in}\caption{\footnotesize{Block diagram of a QIM receiver}}
\label{fig:qim_rx} \vspace{-14pt} \end{figure}

\subsection{QIM Receiver} \label{sec:decode}

Fig.~\ref{fig:qim_rx} shows the block diagram of a QIM receiver. A QIM receiver is based on 
standard RF architecture and is similar to a commodity receiver in terms of complexity and power consumption. 
The QIM decoder is implemented in digital baseband and we simply apply the
same QIM embedding function to the received signal for both $0$ and $1$ bit messages to obtain two 
quantized signals $q_{0}$ and $q_{1}$. These are passed through a minimum distance decoder to compare with the original
received signal $y$ to obtain estimated/received message,

\begin{equation} \label{decode} \hat{m} = \arg\min_{m} dist(y, y_m)
\end{equation}

\section{QIM Techniques} \label{sec:tech} In this section we describe
different QIM techniques that can be used to embed messages in a host
signal.

\subsection{Scalar QIM}

Scalar QIM is the primary QIM technique which was described in the
previous section and can be applied to a real valued or scalar signal.

\subsection{Distortion Compensated QIM } Distortion compensated QIM (DC-QIM) is an extended version of the QIM method that reduces the distortion in the host signal and significantly improves the distortion to robustness trade-off~\cite{qim}. The robustness of QIM is a function of the distance between the quantization points. If we increase the distance, robustness of QIM would also increase. This operation is the same as scaling the QIM embedded signal by a factor, $\alpha$. For example, if a sample point $s_i$ is shifted to a point, $x$, then by scaling it by $\alpha$, the sample point $s_i$ would now be at $\nicefrac{x}{\alpha}$. However, this increases the distortion of the host signal by a factor $\nicefrac{1}{\alpha^2}$ and results in a $\nicefrac{\Delta^2}{12\alpha^2}$ mean distortion. We compensate for the distortion by adding back a fraction, $1-\alpha$ of the host signal. This operation can be represented by the following QIM embedding function,

\begin{equation} \label{eq:dc} \vspace{-2pt}
Q_m(\alpha s) = Q(\alpha s -
d_m    ) + (1-\alpha)s + d_m 
\vspace{-2pt}
\end{equation}

The parameter $\alpha$ is defined as $0 \leq \alpha \leq 1$ where 1 represents the original QIM method. Since, $\alpha$ determines the distortion, a component of noise in the composite signal due to embedding, we can determine the optimal value of $\alpha$ maximizing the signal to noise ratio $SNR$.

\begin{equation} \label{eq:alpha} \vspace{-2pt}
\begin{aligned} SNR(\alpha) &=
\frac{d_1^2}{(1-\alpha)^2 + \alpha^2\sigma_n^2} \\ \\
\frac{\partial SNR(\alpha)}{\partial \alpha} &= 0 \text{ gives } \alpha^* = \frac{D_s}{D_s + \sigma_n^2}, \end{aligned} 
\vspace{-2pt}
\end{equation}

where, $\sigma_n^2$ is the noise power. Hence, the optimal factor
$\alpha^*$ is a function of distortion and noise.

\begin{figure}[t] \vspace{-20pt} \centering \includegraphics[width =
0.95\columnwidth]{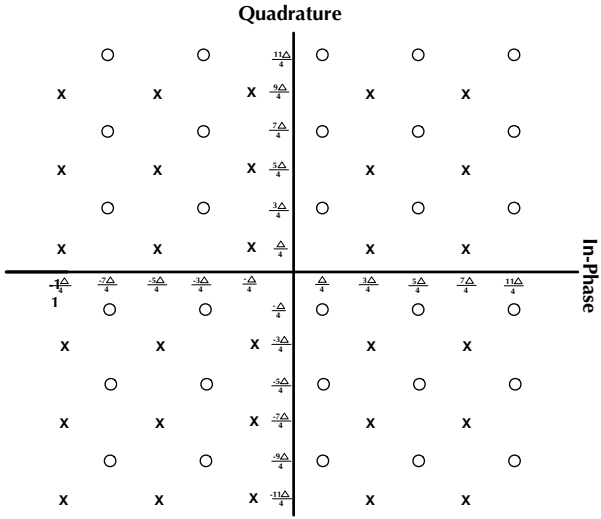} \caption{Constellation diagram for
Lattice QIM with a grid representation. The Xs are quantization points for 1-bit and Os for a 0-bit.} \label{fig:l2} \vspace{-20pt} \end{figure}

\subsection{Lattice QIM}

The QIM technique can be extended to higher dimensional signals by using
a lattice form~\cite{qim}. We can arrange the quantization points to be an integer lattice, $Z^N$, in an N-dimensional Euclidean space $\textbf{R}^N$.
Let's consider a simplified case of a complex or two-dimensional signal
which has both in-phase and quadrature phase components. For N=2, we
will have two grids to quantize to either a 0-bit or 1-bit
representation. As an extension of scalar QIM, the quantization points
for a 0-bit and 1-bit can be represented as co-sets of the lattice,

\begin{equation} \label{l2} \vspace{-2pt}
\Lambda_0 =
\big(\frac{-\Delta}{4}-j\frac{\Delta}{4}\big) + n_i \text{ and }
\Lambda_1 = \big(\frac{\Delta}{4}+j\frac{\Delta}{4}\big) + n_i
\vspace{-2pt}
\end{equation}

where $n_i = 1, 2,..., l$ and $l$ represents the total number of
quantization levels. Fig.~\ref{fig:l2} illustrates the quantization
grid for $N=2$. In lattice QIM, the quantization levels are separated by
$\nicefrac{\Delta}{2}$ distance in in-phase and quadrature components
resulting in an overall distance of $\nicefrac{\Delta}{\sqrt{2}}$.
However, the mean error due to embedding is still $\frac{\Delta^2}{12}$
and because of larger distance between data points, lattice QIM should
perform better than scalar QIM. To implement lattice QIM for complex
wireless signal, we quantize both the real and imaginary parts of the
signal. We use the same embedding function and modify the dither as
\vspace{-2pt}
\begin{equation}
\label{l2_dither} \vspace{-10pt} \small
d_1=\frac{\Delta}{4}+j\frac{\Delta}{4}  \text{ and } d_0=\begin{cases} d_1+
(\frac{\Delta}{2}+j\frac{\Delta}{2}), & \text{if } d_1 \leq 0\\ d_1-
(\frac{\Delta}{2}+j\frac{\Delta}{2}), & \text{otherwise} \end{cases}
\vspace{8pt}
\end{equation}

Finally, we note that the distortion compensation technique described
above can be applied to the real and complex values of the signal to
implement distortion compensated lattice QIM.

\section{Evaluating the QIM Wireless Link} \label{sec:theory}

Evaluation of an embedded communication system (such as
QIM) significantly differs from conventional communication systems.
Traditional systems deal with only one signal whereas
embedded communication systems operate on two signals: the host signal
and the embedded signal. We need to evaluate the impact of QIM on the performance of both the host and the embedded message signal: what is the rate and robustness of the embedded signal, and how much did the embedded process distort the host signal and its impact on the output of the host (legacy) receiver.

\subsection{Performance of Embedded Message}

We start by defining the capacity for QIM embedded message in a Gaussian channel. The output
of the channel is the sum of the input host signal and Gaussian white
noise. Let $\sigma_s^2$ be the average power of the host signal and channel 
noise follows a Gaussian distribution with variance
$\sigma_n^2$. We can express information capacity, i.e., the
supremum of the achievable rate for a real value one dimensional
signal as~\cite{shannon}

\begin{equation}\label{eq:hostcapacity}
C_{host}=\frac{1}{2}\log_2\left(1+\frac{\sigma_s^2}{\sigma_n^2}\right)
\end{equation}

where $\sigma_s^2/\sigma_n^2$ denotes the signal-noise ratio (SNR) and
assumes that the input also follows Gaussian distribution. Since
embedded QIM signal is distortion in the host signal, we can
re-write the capacity expression in terms of distortion by treating the
embedded QIM signal as a form of power-limited communication over a
Gaussian channel:

\begin{equation} \label{eq:capacity}
C_{qim}=\frac{1}{2}\log_2\left(1+\frac{D_s}{\sigma^2_n}\right)
\mathrm{bps/sec/Hz}, \end{equation}

where the distortion constraint is given by Eq.~\ref{eq:distortion}. The
capacity equation, $C_{qim}$, shows that the performance of the embedded message
is directly proportional to distortion experienced by the host signal. The QIM
technique maximizes the capacity of the embedded message for a given distortion of the host signal.
The capacity is also a function of the bandwidth of the host signal and a higher
bandwidth host signal would enable a higher data rate embedded message.
Finally, we note that the capacity in Eq.~\ref{eq:capacity} is the
maximum achievable data rate per unit bandwidth with arbitrarily small
error probability. Practical QIM implementation would require error
correction coding mechanisms to achieve performance close to the limits
promised by channel capacity.

\subsection{Impact on the Host Signal}

The host signal experiences distortion due to perturbations introduced
by the embedded signal which was described in Eq. ~\ref{eq:distortion}.
For a fair comparison across host signals, we introduce
normalized distortion which is independent of the signal strength of the
host signal and is defined as follows,

\begin{equation}\label{eq:dist_norm} D_{s}^{n} = \frac{\sum_{i=1}^{N}
|s_i - x_i|^2}{\sum_{i=1}^{N} |s_i|^2} * 100\% \end{equation}

Finally, in addition to computing the distortion, we will also analyze
the quality of the multimedia signal at the output of the host receiver
to ensure that QIM embedding operation has minimal
impact on the performance of the host signal.

\section{Simulation Results} \label{sec:results} The Wireless QIM
technique is independent of the host signal and is universally
applicable. Here we consider three host signals: AM, FM and
broadcast TV which are ubiquitous in cities. We start with a short primer on the host signals.
%
\vskip 0.05in\noindent{\it {\bf TV.}} In the United States, Digital TV
(DTV) operates in the UHF band from 470-614~MHz with 6~MHz wide channels
and follows the Advanced Television System Committee (ATSC)
standard~\cite{atsc}. ATSC uses 8-level vestigial sideband (8-VSB)
modulation to transmit data. 8-VSB is a digital modulation technique
which uses eight amplitude levels to represent symbols on a 6~MHz
channel. Transmissions from a TV tower can be typically received up to
50 miles.

\vskip 0.05in\noindent{\it {\bf FM Radio.}} FM radio operates in the
87.8-108~MHz frequency band with 200~kHz wide channels. FM uses analog
frequency modulation to encode audio and data i.e. information is
transmitted by varying the frequency of the transmitted RF signal. Most
FM stations can be heard up to 100 miles from the transmit tower.

\vskip 0.05in\noindent{\it {\bf AM Radio.}} In the United States, AM
radio operates in the 525-1705~kHz band with 10~kHz channel
spacing. AM radio uses amplitude modulation to encode data i.e.
information is represented in the amplitude of the signal. AM
signals propagate long distances and have been reported
to have been received 200 miles away from the station.


We evaluate wireless QIM on recorded AM, FM, and TV
signals. The USRP X300~\cite{usrp} was used to record
TV signals centered at 539~MHz (UHF channel 25) with 6.25~MHz sampling
rate and FM signal centered at 106.1~MHz with 200kHz sampling rate. We
use a WebSDR~\cite{websdr} to record an AM signal centered at 1630 kHz
with 8kHz sampling rate. We implement QIM methods described in
Section~\ref{sec:methods} and simulate different channel conditions and data rates using MATLAB.

We embed pseudo-random message bits and implement scalar QIM and scalar
DC-QIM for real valued AM and FM radio signals. For complex TV signals
in addition to scalar QIM, we also evaluate lattice QIM and lattice
DC-QIM. In DC-QIM, we set $\alpha = 0.7$, the optimal value as per
Eq.~\ref{eq:alpha}. We introduce additive white Gaussian noise to
simulate different channel conditions. Finally, a QIM receiver recovers
the transmitted message bits using the algorithm described in
Section~\ref{sec:decode}. The value of $\Delta$, the QIM embedding
parameter is known at the receiver. This is a reasonable assumption
since it can be either pre-set or periodically updated. We evaluate the
system by measuring the impact of QIM on both the host signal and the
performance of the embedded QIM signal for different QIM methods at
different channel conditions and number of quantization levels (affects
distortion).

\subsection{Impact on the Host Signal} \label{sec:imp_host_signal}
The distortion experienced by
the host signal is a function of number of levels $N$ used in the QIM
embedding process. We vary the number of quantization levels from 2 to
45 for embedding random messages in TV, FM and AM host signals and
measure the normalized distortion and its impact on the performance of
the legacy host signal receiver.

\begin{figure}[t] \vspace{-20pt} \centering
\includegraphics[width = 0.8\columnwidth]{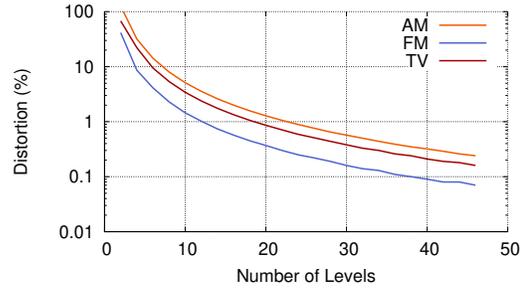} \label{fig:dist_all}
 \vspace{-5pt}
\caption{ Evaluation of distortion for each host signal}
\vspace{-20pt} \label{fig:distortion} \end{figure}

\vskip 0.05in\noindent{\it {\bf Normalized Distortion.}}
Fig.~\ref{fig:distortion} shows the percentage of distortion
experienced by each host signal as a function of number of levels for
scalar DC-QIM technique. The AM signal experiences the most distortion
followed by TV and FM. This is expected since AM uses analog amplitude
modulation to encode data and QIM introduces amplitude perturbations,
which distorts the information carrying amplitude of the AM signal.
Similarly, TV also uses 8 level (digital) amplitude modulation and
amplitude perturbations would impact the digital TV signal but since
digital amplitude modulation is more robust compared to analog
modulation, QIM introduces less distortion in case of TV compared to AM.

The FM signal is the most robust among evaluated signals with
less than 8\% for four quantization levels since amplitude perturbations 
introduced by QIM have minimal impact on the frequency modulated FM signal. 

\begin{figure}[t] \vspace{-5pt} \centering \includegraphics[width =
0.9\columnwidth]{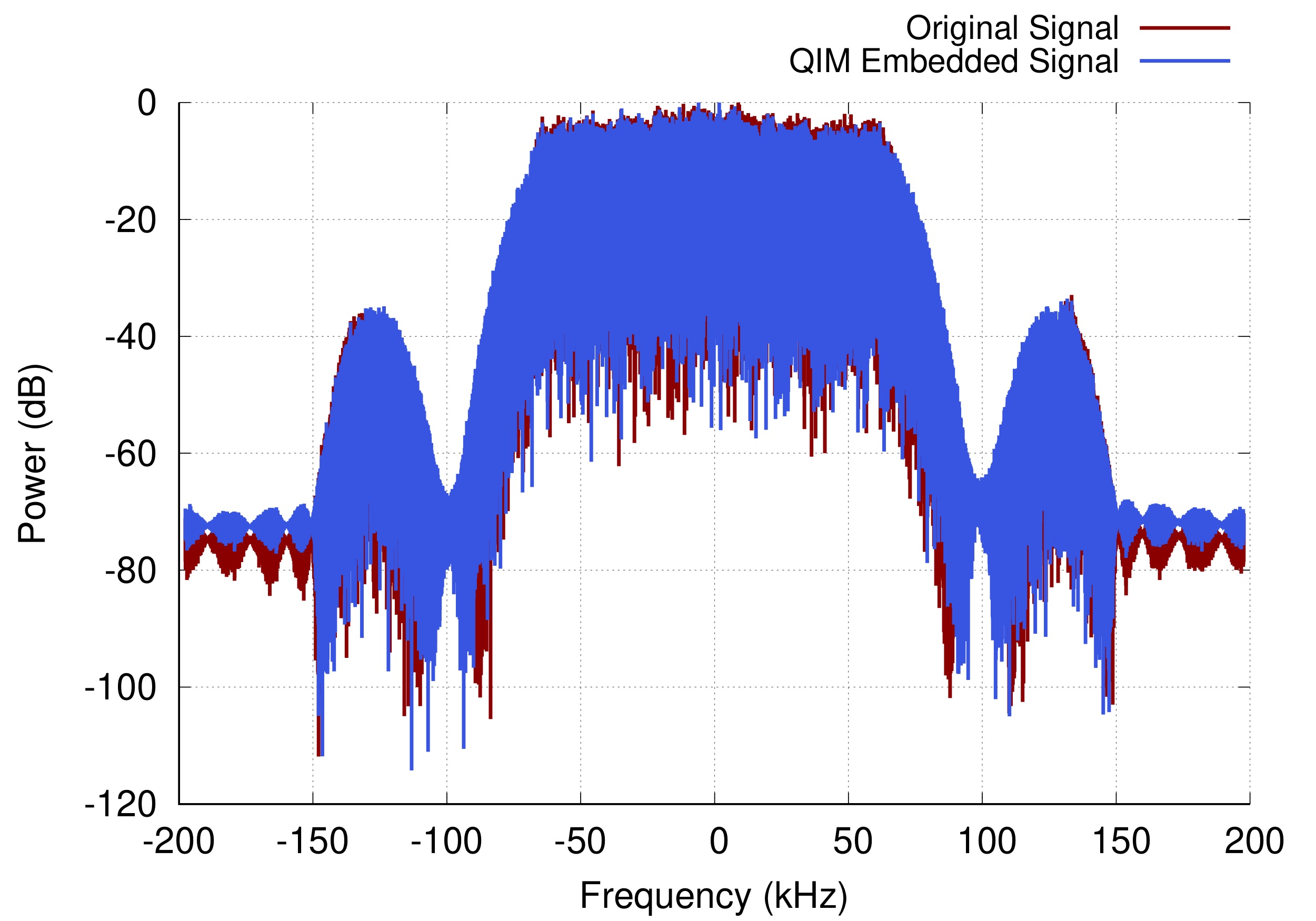} \vspace{-5pt} \caption{ Spectrum of the baseband FM signal before and after QIM embedded data.} \label{fig:spectrum}
\vspace{-15pt} \end{figure}

\vskip -0.0in\noindent{\it {\bf Distortion in the Frequency Domain.}}
Next, we evaluate impact of embedding data using QIM in the frequency domain. Fig.~\ref{fig:spectrum} shows the spectrum of the baseband FM signal before and after the QIM embedding process. The two signals are passed through pulse shaping low pass filters to comply with spectral mask requirements. Our results show that there is small distortion in the in-band spectral characteristics of the baseband signal which corroborate the time domain distortion analysis. The out of band frequency components for both before and after QIM embedded baseband FM signal are atleast 35~dB below the main lobe thereby having minimal impact on any side channels.

\begin{figure*}[t] \vspace{-20pt} \centering
\begin{subfigure}[b]{0.32\textwidth} \includegraphics[width =
\textwidth]{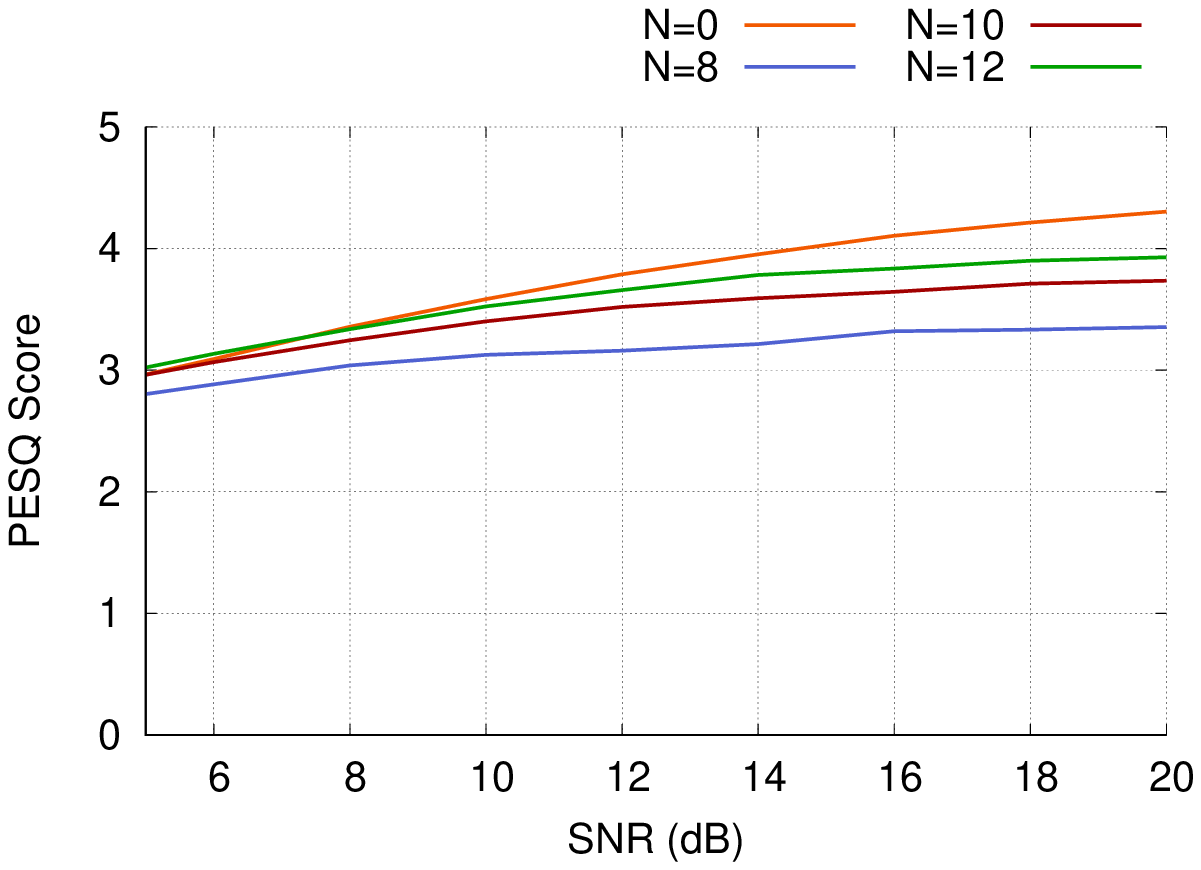} \vspace{-15pt} \caption{\footnotesize Quality of AM
audio.} \label{fig:am_mos} \end{subfigure} ~
\begin{subfigure}[b]{0.32\textwidth} \includegraphics[width =
\textwidth]{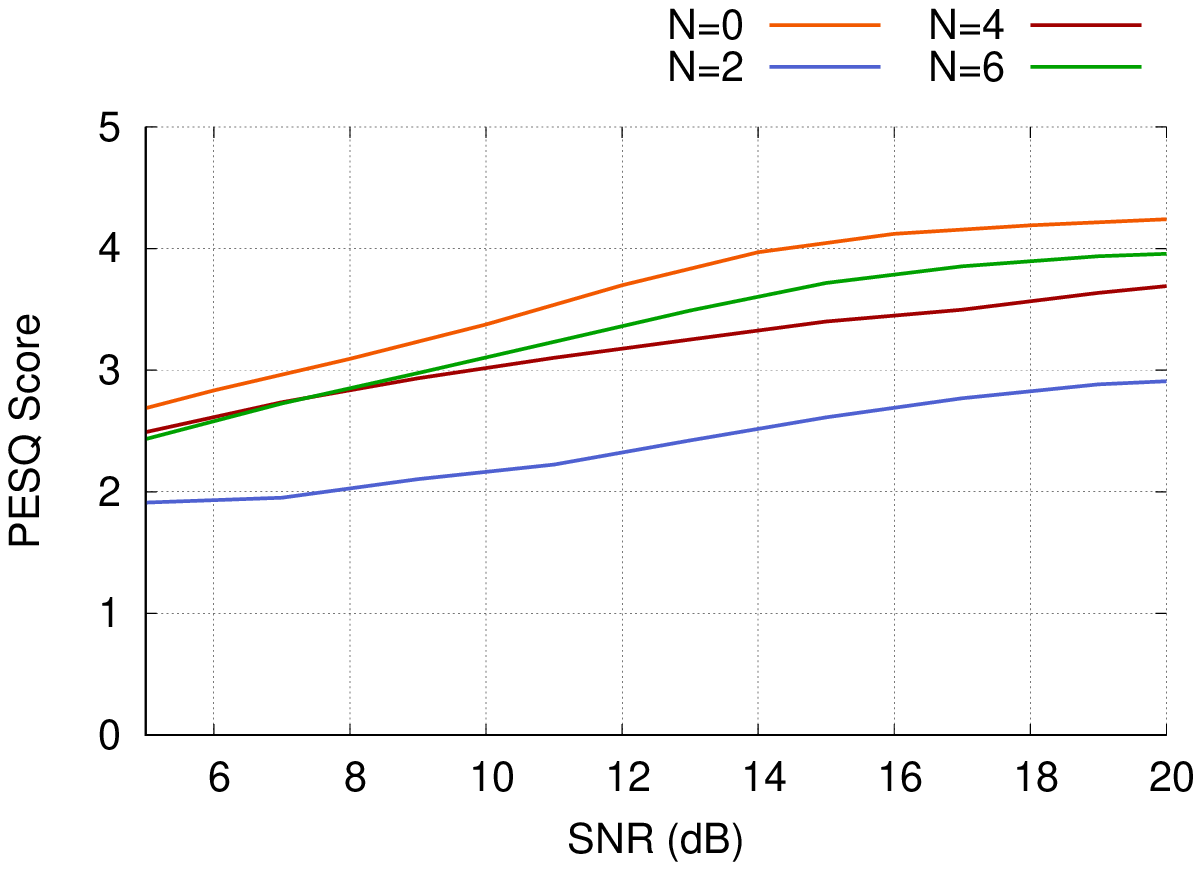} \vspace{-15pt} \caption{\footnotesize Quality of FM
audio.} \label{fig:fm_mos} \end{subfigure} ~
\begin{subfigure}[b]{0.32\textwidth} \includegraphics[width =
\textwidth]{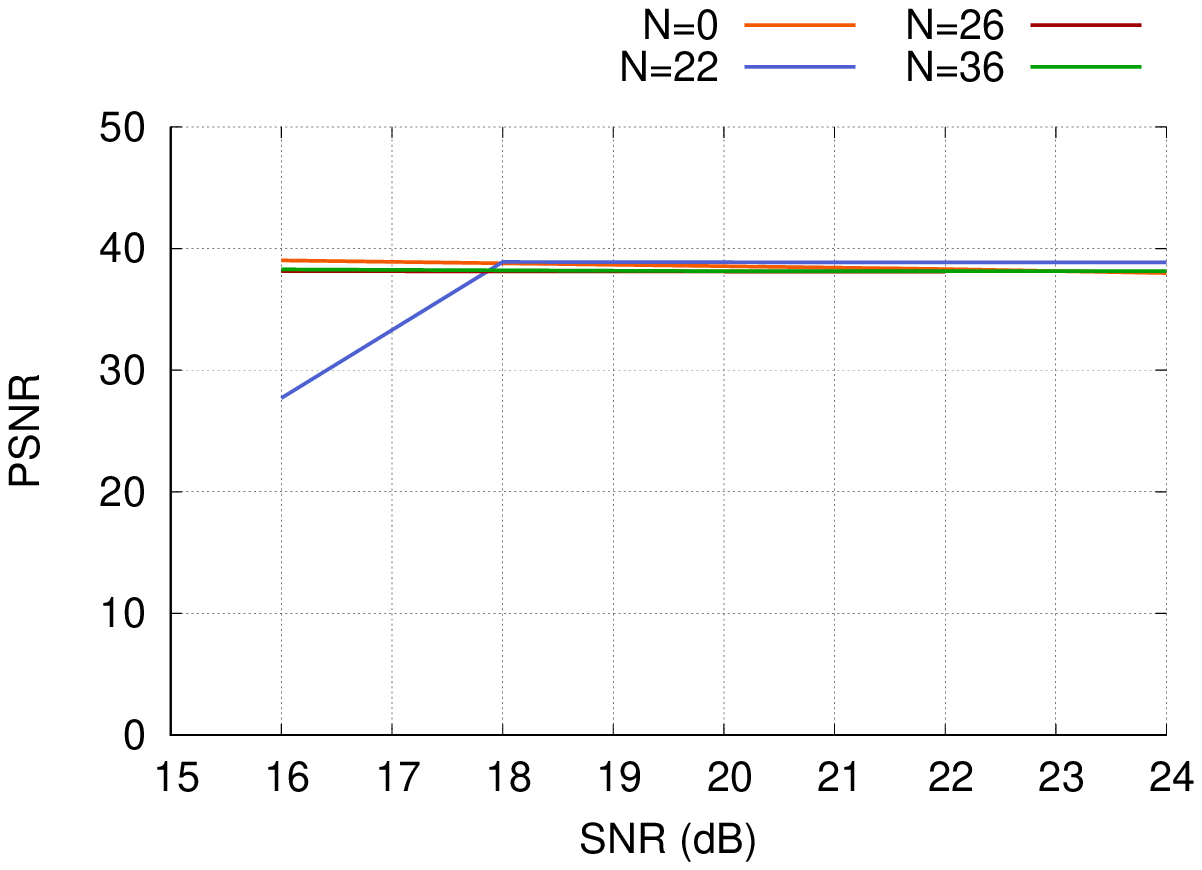} \vspace{-15pt} \caption{\footnotesize Quality of TV
video.} \label{fig:psnr} \end{subfigure} \vspace{-15pt} \caption{Evaluation of the
quality of the host signal multimedia after QIM embedding.}
\vspace{-10pt} \label{fig:mos} \end{figure*}

\vskip 0.05in\noindent{\it {\bf Impact on Host Signal Multimedia.}} 
The next step is to translate the distortion to the quality of
the multimedia audio and video signal carried by AM, FM and TV signals.
This is the key to understanding how QIM impacts the information carried
by the host signal. We use the Perceptual Evaluation of Speech Quality (PESQ) metric to
quantify the quality of demodulated audio from AM and FM signals. The results model
a mean opinion score (MOS) that ranks the quality of speech from 1(bad)
to 5(excellent). As a reference, a PESQ $\geq$ 1 is sufficient for human
hearing~\cite{speech}. We evaluate the PESQ as function of the SNR of
the host signal and distortion introduced by scalar DC-QIM technique.
Fig.~\ref{fig:mos}(a) shows the
audio quality of an AM signal as a function of SNR of the host signal
and the number of quantization levels. We can see that even at the
lowest SNR of 6~dB, PESQ is greater than 2.8 which is sufficient for
most applications. As we increase the SNR and number of quantization
levels, the audio quality improves. The results confirm that QIM has
minimal impact on the audio quality of AM signals. We perform similar
analysis for FM signal and Fig.~\ref{fig:mos}(b) shows the audio quality
of a demodulated QIM embedded FM signal at different host signal SNR and
quantization levels. Since the FM signal is more robust to QIM, we
evaluate the scalar DC-QIM technique for 2-6 quantization levels (42$\%$-2$\%$ distortion) in the FM baseband. However, due
to the robustness of frequency modulation, even at the worst-case SNR of
6~dB and 42$\%$ distortion, PESQ is close to 2, which is satisfactory.

Finally, we evaluate the video quality of the QIM embedded TV signal,
using the peak SNR (PSNR) metric which is the ratio of maximum possible
power of a signal to the power of the distorting noise~\cite{psnr}. The
PSNR is computed as follows 

\begin{equation} PSNR = 10log_{10}(\frac{(max(s))^2}{D_s}) \end{equation}

where $D_s$ is distortion defined in Eq.~\ref{eq:distortion}. For
typical applications, PSNR values between 20-25~dB are acceptable for
wireless systems~\cite{psnr}.

To evaluate the PSNR of the video signal, we first extract the video by
demodulating the QIM embedded TV host signal. We use a software defined
radio (USRP X300) to re-transmit the QIM embedded TV signal at different
SNR to a TV tuner card by Hauppauge to recover the video. The TV signal
was embedded with 20-36 quantization levels which translates to
0.9\%--0.3\% distortion in the TV baseband signal. In
Fig.~\ref{fig:mos}(c) we plot the PSNR of the video output of the TV
tuner card as a function of SNR of the host TV signal and number of
levels. The PSNR of the recovered video was around 34 for majority of
the cases expect for the lowest SNR of 16~dB at the highest distortion.
However, even the lowest values of 28~dB PSNR is acceptable for most
applications. Our analysis considers TV signals above an SNR of 16~dB, a
constraint placed by the sensitivity of the TV tuner card. The TV tuner
was only able to play video from original distortion free TV signal
above an SNR of 16~dB which placed the limit on the SNR of the TV
signal evaluated in this work.

To give readers an intuition about the quality metric used in the
evaluation, we created a composite video of audio and video clips for
AM, FM and TV host signals for different SNR and distortion in the host
signal which can be found at the following web link:

\begin{center} {\color{blue} https://youtu.be/gKn09ctlFMA}. \end{center}


\begin{figure}[t] \vspace{-5pt} \centering \includegraphics[width =
0.9\columnwidth]{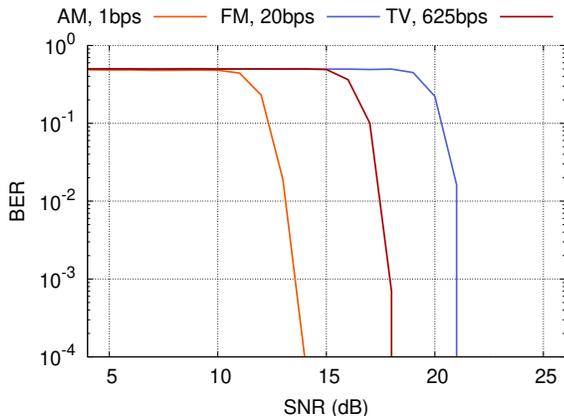} \vspace{-8pt} \caption{A comparison of
performance of all three host signals.} \label{fig:ber_compare_all}
\vspace{-20pt} \end{figure}

\begin{figure*}[t] \vspace{-10pt} \centering
\begin{subfigure}[b]{0.32\textwidth} \includegraphics[width =
\textwidth]{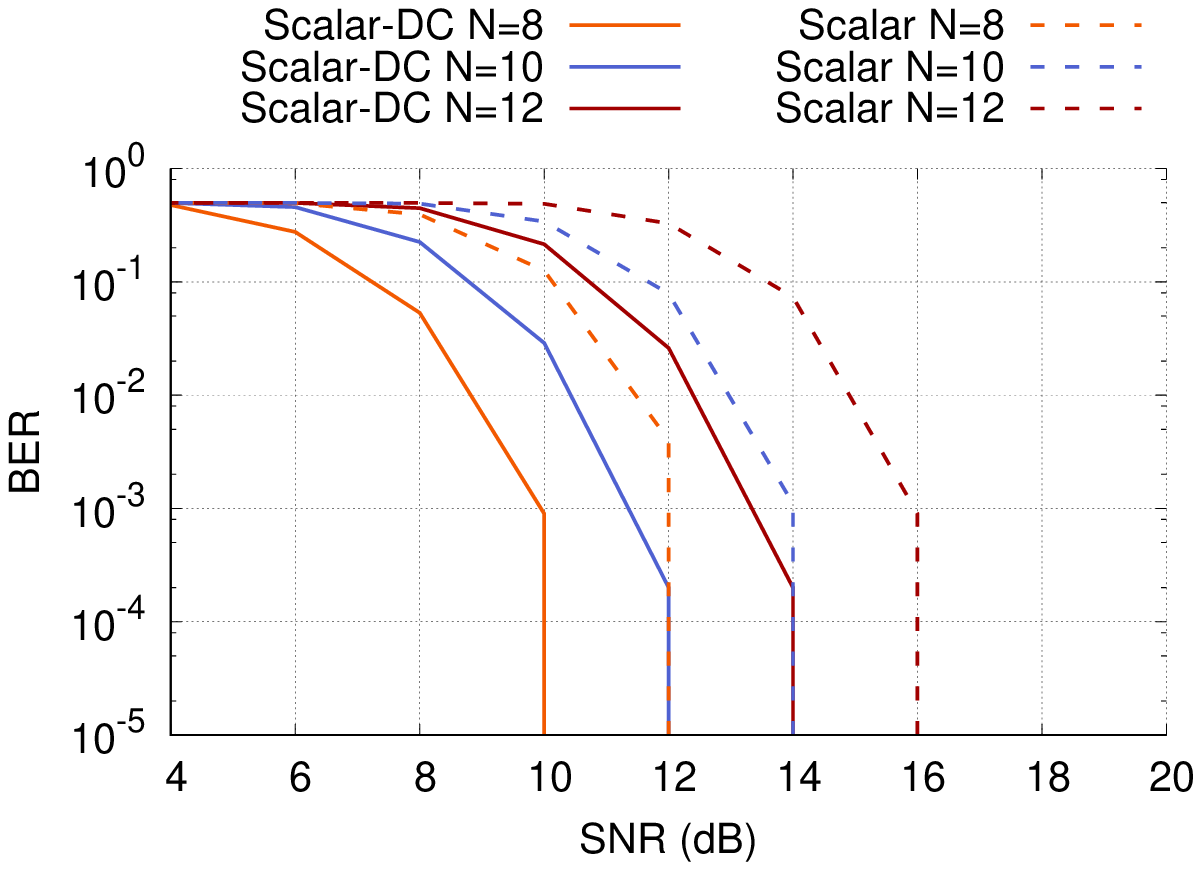} \vspace{-15pt} \caption{\footnotesize Scalar
and Scalar-DC QIM on AM signal.} \label{fig:am_ber} \end{subfigure} ~
\begin{subfigure}[b]{0.32\textwidth} \includegraphics[width =
\textwidth]{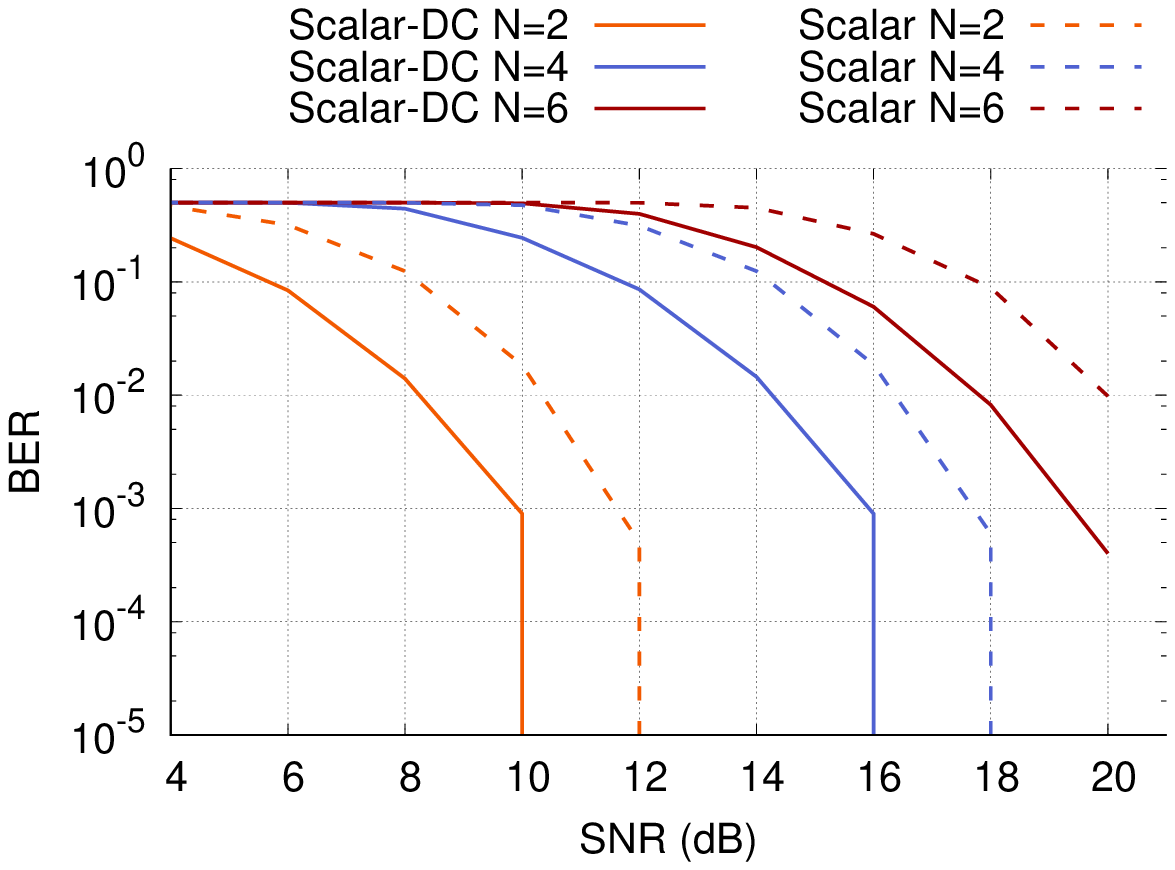} \vspace{-15pt} \caption{\footnotesize Scalar
and Scalar-DC QIM on FM signal.} \label{fig:fm_ber} \end{subfigure} ~
\begin{subfigure}[b]{0.32\textwidth} \includegraphics[width =
\textwidth]{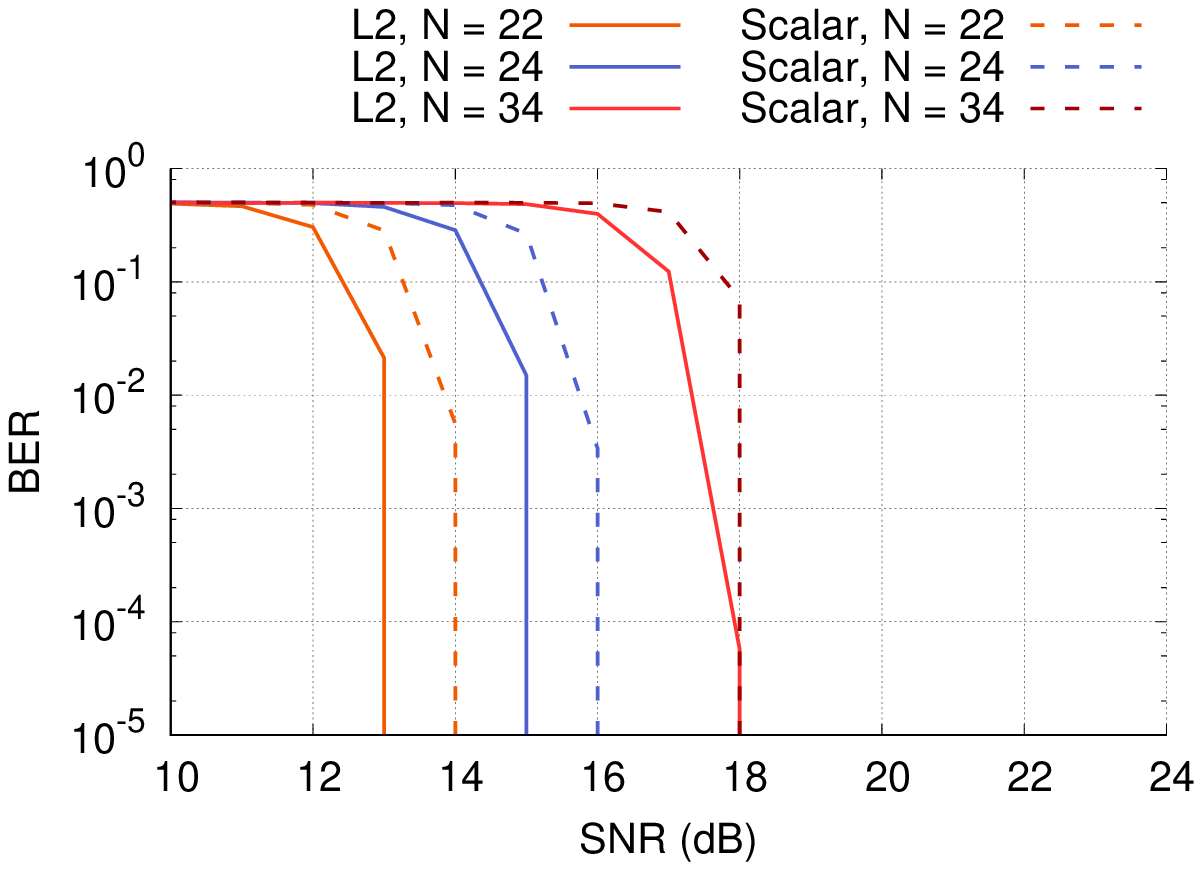} \vspace{-15pt} \caption{\footnotesize Scalar-DC and
Lattice-DC QIM on TV signal.} \label{fig:tv_ber} \end{subfigure}
\vspace{-15pt} \caption{We evaluate the performance of embedding message using QIM on
each of the three host signals.} \label{fig:ber} \vspace{-20pt}
\end{figure*}

\subsection{Performance of the QIM Embedded Message}
\label{sec:results_qim_embedded}

The next step is to evaluate the performance of the QIM embedded message
signal. We only consider scenarios where the distortion and impact on the host multimedia is within acceptable bounds 
described in Section ~\ref{sec:imp_host_signal}.

In Section~\ref{sec:theory} we showed that the information carrying
capacity of QIM, like any communication system, is directly proportional
to the bandwidth of the host signal. To have a fair comparison across
different signals, we normalize the embedded message data rate 
to the bandwidth of the host signal and evaluate performance
for all three host signals. Specifically, we embed at the rate of 1~bps
for 10~kHz bandwidth AM signal, 20~bps for 200~kHz bandwidth FM signal
and 625~bps for 6.25~MHz bandwidth TV signal.
Fig.~\ref{fig:ber_compare_all} shows the BER of the embedded message
using 22 level scalar DC-QIM for the three host signals which translates
to 1\%, 0.3\% and 0.7\% distortion respectively in the AM, FM and TV
signals. We can see that there is a 4~dB difference between AM and TV
and a 7~dB difference between AM and FM. This can be attributed to the
fact that for the same number of levels, the AM signal experiences
0.35\% more distortion compared to TV and 0.76\% more distortion
compared to FM. Since distortion is the embedded signal, a higher
distortion translates to higher signal strength for the embedded message
signal and better performance. We empirically note an
approximate 1~dB increase in performance for every 0.1$\%$ increase in
distortion.

Next, we individually analyze each of the three host signals 
by evaluating the BER of embedded message as a function
of the SNR of the host signal and the signal strength of the embedded
message. We evaluate the AM signal for
8-16 levels at 200bps embedded message rate for both scalar QIM and
scalar DC QIM. Fig.~\ref{fig:ber}(a) shows that the BER decreases with
decrease in number of levels which translates to an increase in
distortion of the host signal or signal strength of the embedded signal.
Specifically, for every decrease in two quantization levels, there is a
2~dB increase in the performance. Finally, the
distortion compensation technique improves the performance by about 2~dB
and this is true for all host signals and both scalar and lattice DC-QIM
methods.

For the distortion tolerant FM signals we embed message at 20~kbps and use 2--6
quantization levels which translates to a distortion of 42\% to 4\%.
Fig.~\ref{fig:ber}(b) shows the performance of scalar and scalar DC QIM
for the host FM signal. We observe a 2~dB increase in performance for
every unit decrease in number of quantization levels.

Finally, we evaluate the TV signal for both scalar and lattice DC-QIM
techniques. Since TV signals are susceptible to distortion, we 
embed messages at 250~bps and use 22--48 quantization levels 
which translates to a distortion of 0.7\% to
0.1\% in the host TV signal. Fig.~\ref{fig:ber}(c) plots the performance
of an embedded QIM message in a host TV signal and we can see that
performance increase with decrease in number of levels or increase in
distortion. Additionally, lattice DC-QIM outperforms scalar DC-QIM by
about 2~dB.

\begin{figure}[t] \vspace{3pt} \begin{subfigure}[b]{\columnwidth}
\centering \includegraphics[width = 0.9\columnwidth]{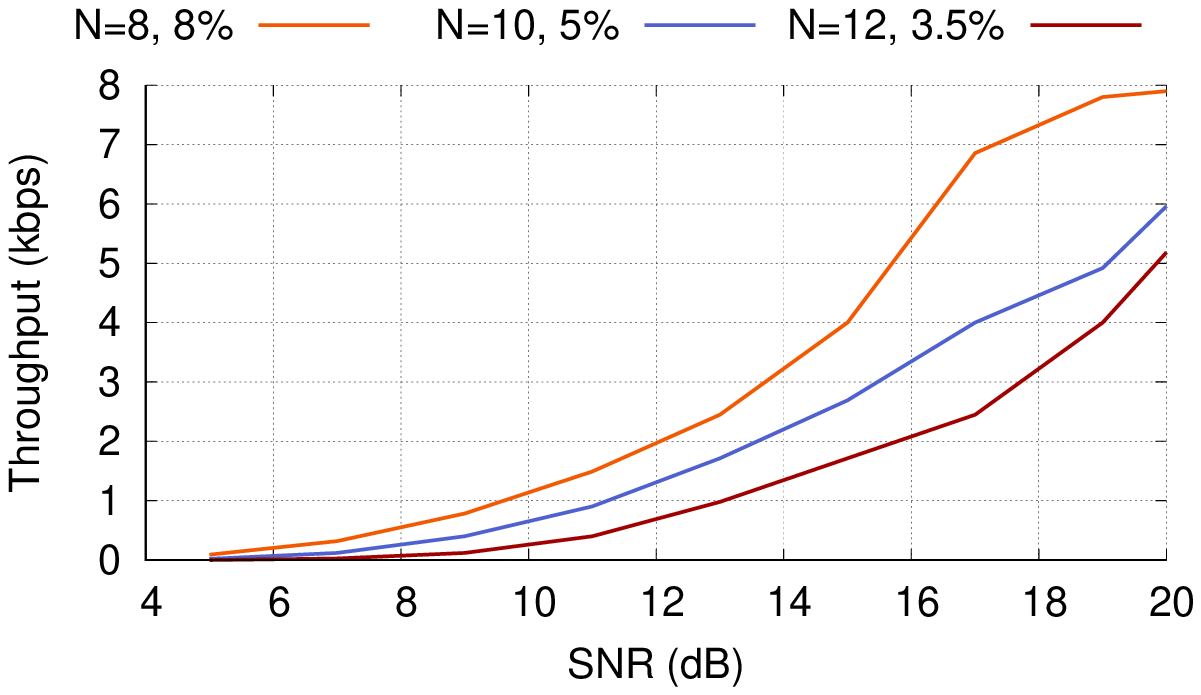}
\vspace{-8pt}
\caption{\footnotesize Achievable throughput with low distortion in host
signal (AM signal).} \label{fig:am_bps} \end{subfigure} ~
\begin{subfigure}[b]{\columnwidth} \centering \includegraphics[width =
0.9\columnwidth]{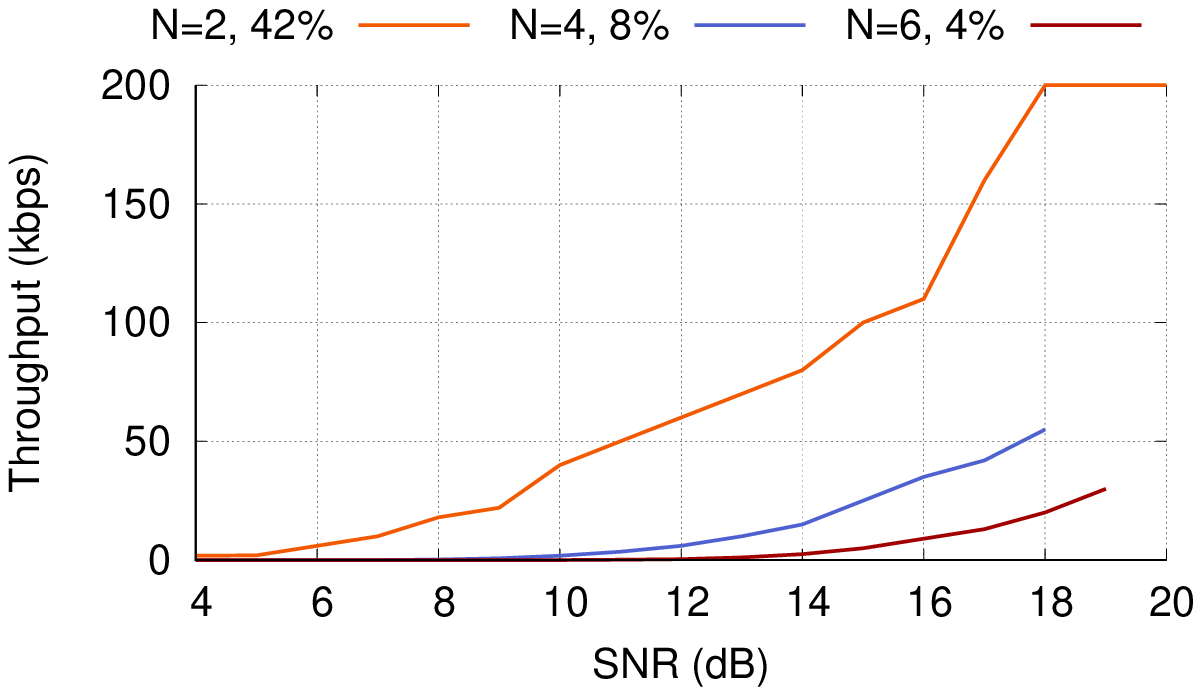}
\vspace{-8pt}
\caption{\footnotesize Achievable throughput with high distortion in
host signal (FM signal).} \label{fig:fm_bps} \end{subfigure}
\vspace{-15pt}
\caption{We evaluate the achievable throughput using Scalar-DC QIM with
high and low distortion in host signals.} \vspace{-20pt} \label{fig:bps}
\end{figure}

\subsection{Achievable Throughput}

We analyze achievable throughput for high and
low distortion in the host signal. For low distortion in a host signal, we consider the 10~kHz wide AM
radio signal and embed messages using scalar DC-QIM method at
different data rates using 8--16 quantization levels (8\%--3.5\%
distortion). Fig.~\ref{fig:bps}(a) shows the throughput as a function of
SNR for different quantization levels. We can see that QIM embedded
message achieves the maximum throughput of 8~kbps for SNR of 20~dB for 8
quantization levels (8\% distortion). Every decrement in two
quantization levels increases the throughput by a factor of 1.5 and
every 2 dB increase in SNR increase the throughput by an average factor
of 3. We can scale these results for FM and TV signals by respectively
multiplying the AM data rates by 20 and 625. For a high distortion case, we consider the 200~kHz wide FM radio signal
and embed messages using scalar DC-QIM method with 2--6 quantization
levels (42\%--4\% distortion). We plot throughput as a function of SNR
for different quantization levels in Fig.~\ref{fig:bps}(b). We achieve a
maximum throughput of 200~kbps. 

\section{Related Work}

Our work is related to recent efforts in spread spectrum watermarking of RF signals~\cite{ssp}. Spread spectrum is a promising technique because it is robust against interfering noise. However, it is a linear method and is susceptible to host signal interference. On the other hand, QIM is a non-linear techniques which is aware of the host signal and therefore efficiently manages host signal interference, resulting in better overall performance. Wireless QIM is also related to inter-protocol communication techniques which enable communication between IoT devices using different wireless standards. This is especially beneficial in the crowded 2.4GHz spectrum where devices with a software modification can using existing hardware to communicate between different devices employing different standards. For instance, the WiZip system uses the presence and absence of packets to encode information for transmission from a Wi-Fi device to a ZigBee device~\cite{wizig}. FreeBee uses variance in timing of regular Wi-Fi beacons to transmit information to ZigBee devices~\cite{freebee}. $B^2W^2$ uses presence and absence of
packets to enable BLE to Wi-Fi communication while concurrently
supporting existing Wi-Fi and BLE communication~\cite{b2w2}. 

All of these techniques are promising, but they are limited to low data rates, are short range and spectrally inefficient since Wi-Fi and ZigBee use drastically different bandwidths. Instead, the Wireless QIM technique can achieve data rates up to 8~kbps with only a 10~kHz wide host AM signal which is 52, 470, and 5.5 order of magnitude higher when compared to WiZig, FreeBee, and $B^2W^2$ which use a significantly wider bandwidth (20~MHz) signal.

\section{Conclusion and Future Work}

We have introduced Wireless QIM technique to embed information into existing signals and communicate with smart devices while having negligible impact on the host signal. We have demonstrated communication at up to of 8~kbps for low bandwidth distortion sensitive AM signals and 200~kbps for higher bandwidth, but distortion resilient FM signals. To the best of our knowledge, this is the first work to use QIM technique to embed messages into wireless signals. We believe Wireless QIM presents a new and exciting opportunity for the radio/TV/cellular providers to enable smart cities and IoT applications by reusing their existing infrastructure and deliver additional value with connectivity at zero spectrum overhead. 

In this paper, although we have evaluated data rate and reliability of the embedded message, we haven't explored errors correcting codes. Development of error correcting codes for Wireless QIM to achieve data rates closer to the theoretical capacity of the channel is an exciting avenue for future research. Finally, this work was focused on downlink communication and in the future work we will extend the Wireless QIM technique for uplink communication as well.

\section{Acknowledgements}
This work is supported in part by NSF award CNS-1305072 and a Google faculty research award.




\bibliographystyle{IEEEtran}
\bibliography{main}

%


\end{document}